\documentstyle[12pt,aaspp4]{aastex}

\begin{document}

\title{A Strong Lyman-$\alpha$ Emitter at $z=6.33$ in the Subaru Deep Field \\
       Selected as an $i^\prime$ Dropout\altaffilmark{1}}

\author{
          T. Nagao            \altaffilmark{2, 3},
          Y. Taniguchi        \altaffilmark{3},
          N. Kashikawa        \altaffilmark{4},
          K. Kodaira          \altaffilmark{5},
          N. Kaifu            \altaffilmark{4},
          H. Ando             \altaffilmark{4},
          H. Karoji           \altaffilmark{6},
          M. Ajiki            \altaffilmark{3},
          M. Akiyama          \altaffilmark{6},
          K. Aoki             \altaffilmark{6},
          M. Doi              \altaffilmark{7},
          S. S. Fujita        \altaffilmark{3},
          H. Furusawa         \altaffilmark{6},
          T. Hayashino        \altaffilmark{8},
          F. Iwamuro          \altaffilmark{9},
          M. Iye              \altaffilmark{4},
          N. Kobayashi        \altaffilmark{7},
          T. Kodama           \altaffilmark{4},
          Y. Komiyama         \altaffilmark{6},
          Y. Matsuda          \altaffilmark{4, 8},
          S. Miyazaki         \altaffilmark{6},
          Y. Mizumoto         \altaffilmark{4},
          T. Morokuma         \altaffilmark{7},
          K. Motohara         \altaffilmark{7}, 
          T. Murayama         \altaffilmark{3},
          K. Nariai           \altaffilmark{10},
          K. Ohta             \altaffilmark{9},
          S. Okamura          \altaffilmark{11, 12},
          M. Ouchi            \altaffilmark{4},
          T. Sasaki           \altaffilmark{6},
          Y. Sato             \altaffilmark{4},
          K. Sekiguchi        \altaffilmark{6},
          K. Shimasaku        \altaffilmark{11},
          Y. Shioya           \altaffilmark{3},
          H. Tamura           \altaffilmark{8},
          I. Tanaka           \altaffilmark{3},
          M. Umemura          \altaffilmark{13},
          T. Yamada           \altaffilmark{4},
          N. Yasuda           \altaffilmark{14}, \&
          M. Yoshida          \altaffilmark{11}
}

\altaffiltext{1}{Based on data collected at the
         Subaru Telescope, which is operated by
         the National Astronomical Observatory of Japan.}
\altaffiltext{2}{INAF -- Osservatorio Astrofisico di Arcetri,
         Largo Enrico Fermi 5, 50125 Firenze, Italy;
         tohru@arcetri.astro.it}
\altaffiltext{3}{Astronomical Institute, Graduate School of Science,
         Tohoku University, Aramaki, Aoba, Sendai 980-8578, Japan}
\altaffiltext{4}{National Astronomical Observatory of Japan,
         2-21-1 Osawa, Mitaka, Tokyo 181-8588, Japan}
\altaffiltext{5}{The Graduate University for Advanced Studies,
         Shonan Village, Hayama, Kanagawa 240-0193, Japan}
\altaffiltext{6}{Subaru Telescope, National Astronomical Observatory 
         of Japan, 650 N. A'ohoku Place, Hilo, HI 96720}
\altaffiltext{7}{Institute of Astronomy, Graduate School of Science,
         University of Tokyo, 2-21-1 Osawa, Mitaka, Tokyo 181-0015, Japan}
\altaffiltext{8}{Research Center for Neutrino Science, Graduate School of 
         Science, Tohoku University, Aramaki, Aoba, Sendai 980-8578, Japan}
\altaffiltext{9}{Department of Astronomy,  Graduate School of Science,
         Kyoto University, Kitashirakawa, Sakyo, Kyoto 606-8502, Japan}
\altaffiltext{10}{Department of Physics, Meisei University, 
         2-1-1 Hodokubo, Hino, Tokyo 191-8506, Japan}
\altaffiltext{11}{Department of Astronomy, Graduate School of Science,
         University of Tokyo, Tokyo 113-0033, Japan}
\altaffiltext{12}{Research Center for the Early Universe, Graduate School
         of Science, University of Tokyo, Tokyo 113-0033, Japan}
\altaffiltext{13}{Center for Computational Physics, University of Tsukuba,
         1-1-1 Tennodai, Tsukuba 305-8571, Japan}
\altaffiltext{14}{Institute for Cosmic Ray Research, University of
         Tokyo, Kashiwa 277-8582, Japan}

\begin{abstract}
We report on the discovery of a star-forming galaxy at $z=6.33$
in the Subaru Deep Field. This object is selected as a candidate
of an $i^\prime$-dropout, high-redshift galaxy around $z=6$ because 
of its red $i^\prime - z^\prime$ color in our deep optical 
imaging survey in the Subaru Deep Field.
Our follow up optical spectroscopy reveals that this object is a 
strong Ly$\alpha$ emitter with only very faint ultraviolet continuum.
The rest-frame equivalent width of the detected Ly$\alpha$ emission is
as much as 130 ${\rm \AA}$.
Thus the light detected in our $z^\prime$ image is largely
attributed to the Ly$\alpha$ emission, i.e., $\sim$40\% of the
$z^\prime$-band flux is the strong Ly$\alpha$ emission, giving a 
very red $i^\prime - z^\prime$ color. 
This is consistent with the photometric property of this object
because the narrow-band data obtained with the $NB921$ filter
shows a significant depression, $z^{\prime} - NB921 = -0.54$ mag.
By using the photometric data, we show that some other objects 
among the 48 $i^{\prime}$-dropout high-redshift galaxy candidates
found in the Subaru Deep Field also show a significant $NB921$ depression. 
We briefly discuss the nature of these $NB921$-depressed objects. 
\end{abstract}


\section{INTRODUCTION}

The star formation in the early universe provides important clues on 
the understanding of both galaxy formation and the ionization state
of intergalactic matter (e.g., Loeb \& Barkana 2001).
Currently, two alternative methods are frequently used to find 
very high-$z$ objects; one is to search for strong emission-line 
objects by using narrow-passband filters (e.g., Hu et al. 2002; 
Ouchi et al. 2003; Ajiki et al. 2003; Kodaira et al. 2003;
see for a review, Taniguchi et al. 2003), which
selectively picks up high-$z$ objects with a strong emission line
such as Ly$\alpha$ (Ly$\alpha$ emitters; hereafter LAEs).
The other method is to search for redshifted Lyman-break objects based 
on broad-band color selections (e.g., Steidel et al. 1996a, 1996b; 
1999; Iwata et al. 2003; Stanway, Bunker, \& McMahon 2003; 
Ouchi et al. 2004; Giavalisco et al. 2004).
Since this method relies on the stellar Lyman-break spectral
feature, this method tends to select high-$z$ galaxies with
strong UV continuum emission, i.e., Lyman-break galaxies (LBGs).
Since some of LBGs also show a strong Ly$\alpha$ emission, LAEs may 
be regarded as a subclass of LBGs (see, e.g., Shapley et al. 2003).

And very recently, based on imaging surveys with broad pass-band 
filters, objects with a very red $i^\prime - z^\prime$ color (i.e.,
$i^{\prime}$ dropout objects) are intensively explored as candidates of
galaxies at $z \gtrsim 6$, where the cosmic reionization ended
(e.g., Stanway et al. 2003; Bouwens et al. 2003, 2004;
Dickinson et al. 2004; Bunker et al. 2004).
In this Letter, we report on the discovery of a luminous 
star-forming galaxy at $z=6.33$ in the Subaru Deep Field (SDF).
This object is originally selected as a candidate of 
$i^{\prime}$-dropout high-redshift galaxy because of its red 
$i^\prime - z^\prime$ color by the imaging survey for SDF.
This object is interesting because our follow-up spectroscopy
shows that it presents only very faint UV continuum emission.
We adopt a cosmology with 
($\Omega_{\rm tot}$, $\Omega_{\rm M}$, $\Omega_{\Lambda}$)
=(1.0, 0.3, 0.7) and $H_0$ = 70 km s$^{-1}$ Mpc$^{-1}$ 
throughout this Letter. 
We use the AB photometric system for optical magnitudes.

\section{OBSERVATIONS}

\subsection{Optical Deep Imaging}

We have carried out a very deep optical imaging survey in the SDF
centered at $\alpha$(J2000) = $13^{\rm h} ~ 24^{\rm m} ~ 38\fs9$ 
and $\delta$(J2000) = $+27^\circ ~ 29' ~ 25\farcs9$,
by using Suprime-Cam (Miyazaki et al. 2002) which consists of 
$5\times 2$ CCDs of 2k $\times$ 4k pixels with a pixel scale of 
$0\farcs202$ pixel$^{-1}$, on the 8.2m Subaru Telescope 
(Kaifu et al. 2000; Iye et al. 2004).
The observations were made with five broad pass-band filters,
$B$, $V$, $R_{\rm c}$, $i'$, and $z'$, and two narrow pass-band filters,
$NB816$ and $NB921$. The central wavelengths and the half widths of
the transmittance of the two narrow pass-band filters are
($\lambda_{\rm c}$, $\Delta \lambda_{\rm FWHM}$)
= (8150${\rm \AA}$, 120${\rm \AA}$) and (9196${\rm \AA}$, 132${\rm \AA}$); 
see Ajiki et al. (2003) and Kodaira et al. (2003) for more details.
The data were collected in several observing runs during a period
between 2001 and 2003. 
A summary of the imaging observations is given in Table 1.

The individual CCD data were reduced and
combined using IRAF and the mosaic-CCD data reduction software 
developed by Yagi et al. (2002).
The combined images for individual bands were aligned and
smoothed with Gaussian kernels to match their seeing sizes.
The PSF FWHM of the final images is $0\farcs98$.
Exposure times and limiting magnitudes are listed in Table 1.
Photometric calibrations are made using usual standard stars.
Source detection and photometry are performed using
SExtractor version 2.1.6 (Bertin \& Arnouts 1996).
Here the $z^{\prime}$ image is used for the source detection.
The effective area of this imaging survey is 761 arcmin$^2$.

We select candidates of $i^{\prime}$-dropout galaxies at $z\sim 6$ 
imposing the following four criteria on the 
$z^{\prime}$-selected sample;
$i^\prime - z^\prime > 1.5$,
$z^\prime < 26.1$ (above $5\sigma$),
$B > 28.5$ (below $3\sigma$),  and
$R_{\rm c} > 27.8$ (below $3\sigma$),
where the magnitudes are measured by adopting the aperture size of
a $2\farcs0$ diameter.
We finally obtain a photometric sample of 48 $i^{\prime}$-dropout 
galaxy candidates at $z \sim 6$.

\subsection{Optical Spectroscopy}

In order to investigate the nature of $i^{\prime}$-dropout high-$z$
galaxy candidates found in our
optical imaging survey, we have completed optical spectroscopy of 
9 objects in our 48 $i^{\prime}$-dropout sample up to now. 
We used the Subaru Faint Object Camera And Spectrograph (FOCAS; Kashikawa 
et al. 2002) on 2003 May and June.
Among the 9 objects, we concentrate our discussion on an
interesting object, SDF J132440.6+273607, in this Letter. 
The observational properties of the other $i^{\prime}$-dropout sample
will be described elsewhere.
For the spectroscopy of SDF J132440.6+273607,
the 300 lines mm$^{-1}$ grating was used with an order cut filter O58.
The wavelength coverage was $\sim$ 5800 \AA ~ to 10000 \AA ~
with a pixel resolution of 1.35 \AA.
The use of a $0\farcs8$ slit gave a spectroscopic resolution of
9.0 \AA ~ at 9000 \AA ~ 
that is measured by atmospheric OH airglow lines.
This corresponds to the spectral resolution of $R \simeq 1000$.
The spatial sampling was 0$\farcs$31 per resolution
element as we adopted 3 pixel on-chip binning.
We obtained seven 1800s exposure frames for SDF J132440.6+273607.
Typical seeing size was $0\farcs4 - 0\farcs6$ during the observation.
We also obtained the spectra of
spectroscopic standard stars Hz 44 and Feige 34 for flux calibration.

\section{RESULTS}

As mentioned above, SDF J132440.6+273607 was selected by its red
$i^\prime - z^\prime$ color, i.e., so-called $i^\prime$-dropout selection.
Indeed this object is not detected above the 3$\sigma$ level on
the $B$, $V$, $R_{\rm c}$, $i^\prime$, and $NB816$ images.
SDF J132440.6+273607 is clearly detected in the $z^\prime$ and $NB921$ 
images; the 2$\farcs$0 aperture magnitudes are
$z^\prime = 25.66$ mag and $NB921 = 26.20$ mag. 
Since the 3$\sigma$ limiting magnitude of $i^\prime$ is 
$i^\prime_{\rm lim} = 27.43$ mag,
SDF J132440.6+273607 has a color of $i^\prime - z^\prime > 1.77$,
satisfying our criterion for the $i^\prime$-dropout selection,
$i^\prime - z^\prime > 1.5$.
The optical thumbnail images are shown in Figure 1.
The image size of SDF J132440.6+273607 in FWHM is 5.8 pixels or 
$1\farcs1$ in the $z^\prime$-band image and 6.0 pixels or 
$1\farcs2$ in the $NB921$-band image, respectively.
Both of these two images may be only marginally larger than the
PSF size.

In Figure 2, we show the sky-subtracted optical position-velocity 
spectrogram and sky-subtracted one-dimensional spectrum of 
SDF J132440.6+273607, obtained by FOCAS on Subaru.
The aperture size to extract the one dimensional spectrum
is 5 binned pixels ($1\farcs6$).
We can see a prominent single emission line whose peak is at
8909${\rm \AA}$, and no significant continuum emission in the
whole of the covered wavelength range.
The spectral feature seen at $\sim$8885${\rm \AA}$ is a
residual of the sky subtraction.
The observed velocity width of the emission line is
$\sim8{\rm \AA}$ in FWHM, roughly comparable to the instrumental width.
However, we can see a clear asymmetric feature below the half maximum of
the emission line; i.e., a very prominent redward tail 
which extends up to $\sim8930{\rm \AA}$.
This asymmetry and the photometric property strongly suggest that
the detected emission line is the Ly$\alpha$ emission and
that SDF J132440.6+273607 is at $z = 6.33$.
Note that this object is thought not to be an emission-line object
at lower redshift; i.e., an H$\alpha$ emitter at $z\sim0.36$,
an [O {\sc iii}]$\lambda$5007 emitter at $z\sim0.78$, or
an H$\beta$ emitter at $z\sim0.83$. This is because there is no
emission lines in the covered wavelength range other than
the prominent emission at 8909${\rm \AA}$. 
We cannot completely deny the possibility that SDF J132440.6+273607
is a [O {\sc ii}]$\lambda$3727 emitter at $z\sim1.39$, because
the [O {\sc ii}]$\lambda$3727 doublet cannot be resolved by our 
observation (the expected wavelength separation of the
[O {\sc ii}] doublet is 6.7${\rm \AA}$).
However, this does not seem plausible because
the observed red tail or shoulder of the detected emission line
is hard to be understood if the emission line is 
[O {\sc ii}]$\lambda$3727. Therefore, taking both the photometric and
the spectroscopic properties into account, we conclude that 
SDF J132440.6+273607 is a strong Ly$\alpha$ emitter without
strong UV stellar continuum emission at $z = 6.33$.
The observed Ly$\alpha$ flux is $(4.0 \pm 0.1) \times 10^{-17}$
ergs s$^{-1}$ cm$^{-2}$, where the slit loss is not taken into account.
The Ly$\alpha$ luminosity is thus calculated to be 
1.8 $\times 10^{43}$ ergs s$^{-1}$.

Based on the Ly$\alpha$ flux obtained by the spectroscopy,
we can derive a Ly$\alpha$-corrected $z^\prime$ magnitude, 
$z^\prime_{\rm cor}$ = 26.16 mag, which is very close to
the $NB921$ magnitude.
Since our $i^{\prime}$-dropout search is complete only down to
$z^\prime = 25.93$ mag, we could not pick up SDF J132440.6+273607 
as an $i^{\prime}$-dropout object if there were no strong
Ly$\alpha$ emission; i.e., $i^\prime - z^\prime_{\rm cor} > 1.27$,
which does not satisfy our $i^{\prime}$-dropout criterion 
($i^\prime - z^\prime > 1.5$).
Then we focus on the depressed $NB921$
magnitude compared to the $z^{\prime}$ magnitude; i.e., 
$z^{\prime} - NB921 = -0.54$ mag.
If we assume that the depression of the $NB921$ magnitude is caused
only by the contribution of the Ly$\alpha$ emission in the 
$z^{\prime}$ band, the required flux of the Ly$\alpha$ emission is
$(4.2 \pm 2.2) \times 10^{-17}$ ergs s$^{-1}$ cm$^{-2}$.
This is consistent with the spectroscopically measured Ly$\alpha$ flux,
where we assume that all of the $NB921$ flux is attributed to the
stellar UV continuum emission. This assumption leads to the UV continuum
flux density of $(4.3 \pm 1.0) \times 10^{-20}$ ergs s$^{-1}$ cm$^{-2}$ 
${\rm \AA^{-1}}$, which is consistent with the non-detection of the 
continuum emission in our optical spectroscopy of SDF J132440.6+273607
(Figure 2).

\section{DISCUSSION}

As presented in the last section, our spectroscopic observation
discloses that SDF J132440.6+273607 is a strong LAE
with only faint UV stellar emission, despite the fact that it has been 
originally selected as an $i^{\prime}$-dropout galaxy candidate 
by its red $i^\prime - z^\prime$ color.
This LAE is thought to be an actively star-forming galaxy, though
we cannot completely reject the possibility that it is a narrow-line QSO.
Although the stellar continuum emission is not detected in our
spectroscopy, we can estimate the equivalent width of the Ly$\alpha$ 
emission based on the $NB921$ data. The photometrically determined 
Ly$\alpha$ equivalent width of SDF J132440.6+273607 is 
$EW_{\rm obs}({\rm Ly}\alpha) = 980 {\rm \AA}$. We then obtain the 
equivalent width at the rest frame; $EW_0({\rm Ly}\alpha) = 130 {\rm \AA}$.
The derived equivalent width is very large as a broad-band selected
galaxy, since a very small fraction of LBGs at $z \sim 3$ show
$EW_0({\rm Ly}\alpha) > 100 {\rm \AA}$ (e.g., Steidel et al. 1999;
Shapley et al. 2003). Rather this extremely large equivalent width seems 
more similar to the narrow-band selected galaxies (e.g., 
Malhotra \& Rhoads 2002; Ajiki et al. 2003; Hu et al. 2004).
Hu et al. (2004) recently found a number of LAEs with 
$EW_0({\rm Ly}\alpha) > 100 {\rm \AA}$ at $z \sim 5.7$ by using a 
narrow-passband filter.
Note that LBGs and LAEs do not necessarily have completely distinct
properties such as $EW_0({\rm Ly}\alpha)$, because they are
defined just by their selection method.
However, it is not clear how these two populations overlap or
how different statistically, especially at the very high-$z$ universe
where the number of galaxies that have been investigated by 
spectroscopic observations is very small. Spectroscopic observations 
on a large number of high-$z$ galaxies are also necessary to investigate 
whether or not the $EW_0({\rm Ly}\alpha)$ frequency distribution depends 
on redshift, which is also a very interesting issue.
Forthcoming intensive spectroscopic observations for high-$z$ galaxies
will hopefully solve these problems.

Our finding of SDF J132440.6+273607 naturally leads to the 
following question: is this object a very peculiar object, or
does any photometrically constructed $i^{\prime}$-dropout sample
generally contain a lot of galaxies with only faint UV stellar continuum?
In order to investigate this issue, we focus on the 
photometrically-selected 48 $i^{\prime}$ dropout LBG candidates in SDF.
In Figure 3, we show the $z^{\prime} - NB921$ colors of the 
$i^{\prime}$ dropout LBG candidates as functions of the 
$z^{\prime}$ magnitude and the $i^\prime - z^\prime_{\rm cor}$ color.
The $NB921$ depression of SDF J132440.6+273607
is $z^{\prime} - NB921 = -0.54$ as mentioned above, whose statistical
significance is roughly $\sim$2$\sigma$. Interestingly,
there are other three $i^{\prime}$-dropout galaxy candidates which 
show significant $NB921$ depression above the 3$\sigma$ level
(objects b, c, and e in the Figure 3).

There are three possibilities for such $NB921$-depressed, 
$i^{\prime}$-dropout high-$z$ galaxy candidates. The first possibility 
is the LAE at $6.0 \lesssim z \lesssim 6.5$ which corresponds to 
$8510 {\rm \AA} \lesssim \lambda_{\rm Ly \alpha} \lesssim 9110 {\rm \AA}$,
just similar to the case of SDF J132440.6+273607. In this redshift range, 
the Ly$\alpha$ emission falls only in the $z^{\prime}$ band and not in 
the $NB921$ one, which results in the $NB921$ depression. The second 
possibility is the ``$NB921$-dropout LBG''; i.e., the LBGs at 
$z \gtrsim$ 6.6. In this case, the original categorization as 
$i^{\prime}$ dropout LBGs is not wrong, because the objects show the UV 
stellar continuum with a significant Lyman break. The third possibility 
is the $i^{\prime}$ dropout galaxy with a significant absorption feature 
at the $NB921$ range, although it is not clear whether or not such an 
object really exists in such a high-redshift universe. In order to 
investigate these ideas further, we plot the expected colors of high-$z$ 
galaxies with a strong Ly$\alpha$ emission on the $z^{\prime} - NB921$ 
versus $i^\prime - z^\prime$ two-color diagram (Figure 3). Here we use 
the galaxy evolution model of Bruzual \& Charlot (2003) by assuming that 
the star-formation rate of model galaxies is proportional to 
exp ($-t/\tau$), where $\tau$ = 1 Gyr and $t$ = 1 Gyr are adopted.
Here we examine only three cases, i.e., 
$EW_0({\rm Ly}\alpha) =$ 65, 130, and 260 ${\rm \AA}$. The $NB921$ 
depression of $-0.2$, $-0.6$ and $-1.0$ mag can be realized for the 
cases of $EW_0({\rm Ly}\alpha) =$ 65, 130, and 260 ${\rm \AA}$ for the 
first possibility. On the other hand, more significant $NB921$ depression
appears for the second possibility. As for the second possibility, a very 
red color of $i^\prime - z^\prime > 3$ is also expected. However, an 
extremely deep $i^\prime$ image is required to discriminate these 
possibilities clearly. Since the two $NB921$-depressed $i^{\prime}$ 
dropout objects with $i^\prime - z^\prime \sim 3$ may be $NB921$-dropout 
galaxies, a follow-up spectroscopic observation will be very interesting
to access galaxies at $z \gtrsim 6.6$.

Here we discuss the star-forming activity in SDF J132440.6+273607.
The Ly$\alpha$ flux of this object measured by our spectroscopy
is 4.0 $\times 10^{-17}$ ergs s$^{-1}$ cm$^{-2}$, which
corresponds to the Ly$\alpha$ luminosity of 
1.8 $\times 10^{43}$ ergs s$^{-1}$.
We can estimate the star-formation rate (SFR) by using the
following relation;
$SFR({\rm Ly}\alpha) = 9.1 \times 10^{-43} L({\rm Ly}\alpha)$ 
$M_\odot {\rm yr}^{-1}$ (Kennicutt 1998; Brocklehurst 1971).
We then obtain $SFR({\rm Ly}\alpha) \simeq 16 M_\odot$ yr$^{-1}$.
This is a lower limit because no correction was made for
possible absorption effects on the Ly$\alpha$ emission.
We can also estimate the SFR by the luminosity of the UV stellar
continuum by adopting the following relation;
$SFR({\rm UV}) = 1.4 \times 10^{-28} L_{\nu} M_{\odot} {\rm yr}^{-1}$
(Kennicutt 1998). By using the UV continuum flux density measured
from the $NB921$ magnitude, $f_\nu$ ($NB921$) = $1.2 \times 10^{-30}$ 
ergs s$^{-1}$ cm$^{-2}$ Hz$^{-1}$, we obtain 
$SFR({\rm UV}_{1255}) \simeq 10.4 M_\odot$ yr$^{-1}$.
One of the reason why $SFR({\rm UV}_{1255})$ is smaller than 
$SFR({\rm Ly}\alpha)$ may be that $SFR({\rm UV})$ would be
underpredicted for for objects with a very large $EW_0({\rm Ly}\alpha)$.
This is because such a huge $EW_0({\rm Ly}\alpha)$ is hard to be 
created by a continuous star formation with a normal initial-mass 
function, which is assumed in the above relation between the SFR and
the luminosity of the UV stellar continuum.
Schaerer (2000) presented that $SFR({\rm UV})$ is underestimated
for star-forming galaxies with an age of $< 10^8$ yr.
Taking all of above considerations into account, we conclude that
the SFR of SDF J132440.6+273607 is $> 16 M_\odot$ yr$^{-1}$.
Since there may be a large number of objects with large
$EW_0({\rm Ly}\alpha)$ as suggested by Figure 3, systematic 
spectroscopic studies on $i^{\prime}$-dropout galaxies are
crucial to understand the star-forming activities of galaxies
at $z \gtrsim 6$, which is a very important epoch in relation to
the cosmic reionization.

\vspace{0.5cm}

We thank the referee, A. Bunker, for his valuable comments.
We also thank the Subaru Telescope staff for
their invaluable assistance.
TN and MA are JSPS fellows.


\clearpage


\begin{deluxetable}{ccc}
\tablenum{1}
\footnotesize
\tablecaption{A summary of optical imaging observations}
\tablewidth{0pt}
\tablehead{
 \colhead{Band} &
 \colhead{Exposure Time} &
 \colhead{$m_{\rm AB}$(lim)\tablenotemark{a}} \\
 \colhead{} & 
 \colhead{(s)} &
 \colhead{}
}
\startdata
$B$         & 35700 & 28.45 \\
$V$         & 20400 & 27.74 \\
$R_{\rm c}$ & 36000 & 27.80 \\
$i^{\prime}$& 48060 & 27.43 \\
$z^{\prime}$& 30240 & 26.62 \\
$NB816$     & 36000 & 26.63 \\
$NB921$     & 53940 & 26.54 \\
\enddata
\tablenotetext{a}{{}Limiting magnitude (AB) for a 3$\sigma$ 
                  detection on a $2\farcs0$ diameter aperture.}
\end{deluxetable}
\clearpage


\begin{figure*}
\epsscale{1.00}
\plotone{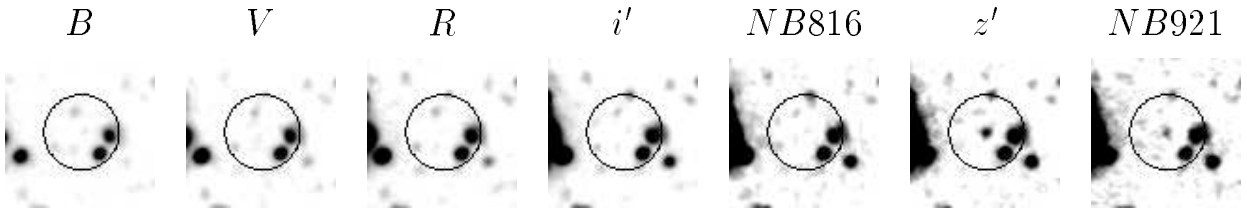}
\caption{
Thumbnail images of SDF J132440.6+273607.
The square regions around SDF J132440.6+273607 in the
$B$, $V$, $R_{\rm c}$, $i^{\prime}$, $NB816$, $z^{\prime}$, and
$NB921$ images are shown from left to right. 
Panel and circle sizes are
16${\arcsec}$ and 8${\arcsec}$, respectively.
\label{fig1}}
\end{figure*}

\begin{figure*}
\epsscale{0.80}
\plotone{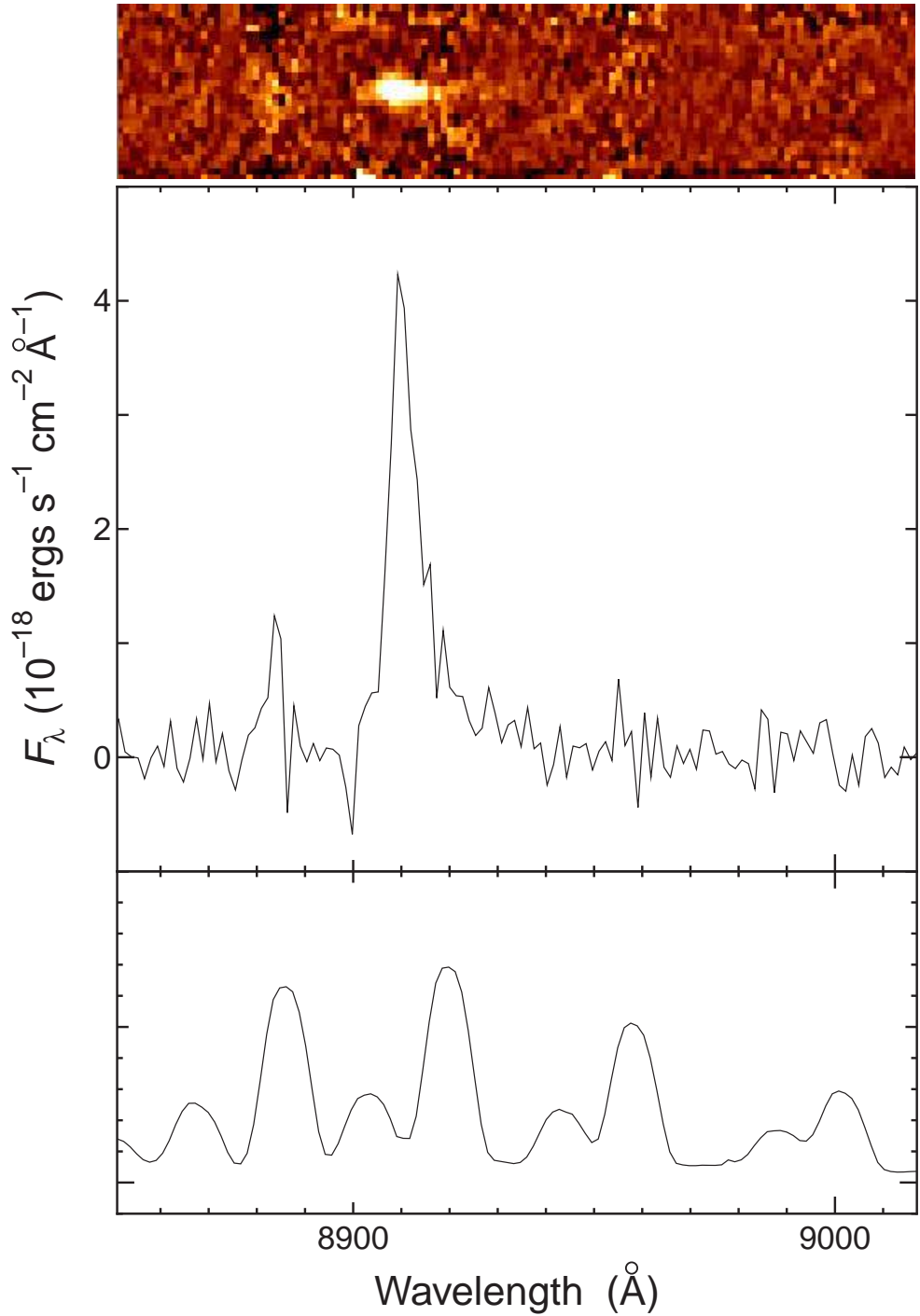}
\caption{
Sky-subtracted optical position-velocity spectrogram
(upper panel) and one-dimensional spectrum (middle panel)
of SDF J132440.6+273607 obtained with FOCAS on Subaru.
Spectrum of the sky emission is also shown in the lower panel.
The spatial size of the displayed area in the upper panel
is 8$\farcs$1.
\label{fig2}}
\end{figure*}

\begin{figure*}
\epsscale{1.00}
\plotone{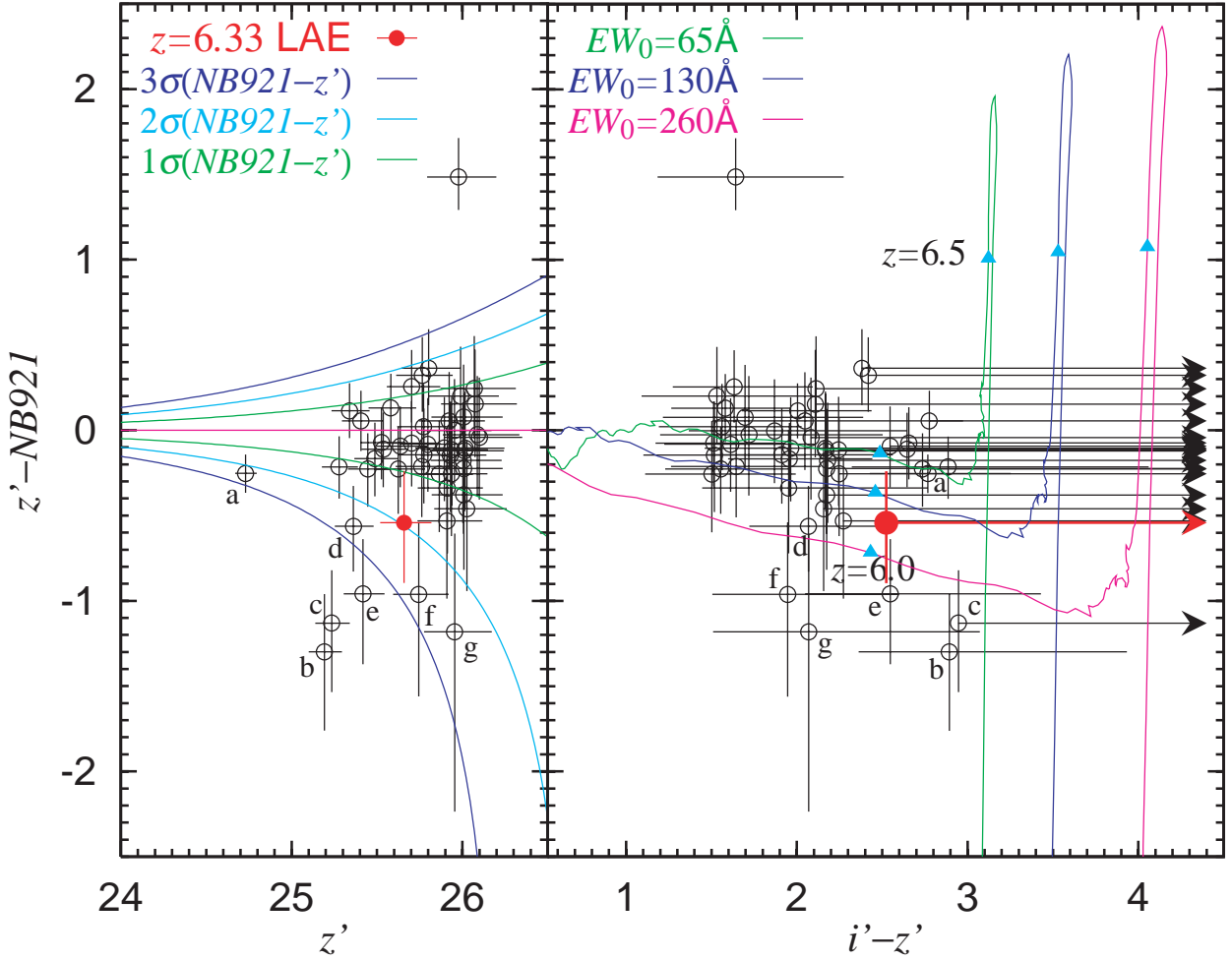}
\caption{
($Left$)
Color of $z^{\prime} - NB921$ of our 48 $i^{\prime}$-dropout
high-$z$ galaxy candidates as a function of the $z^{\prime}$ magnitude.
The error-bars denote 1$\sigma$ uncertainties.
The data of SDF J132440.6+273607 is marked by a red circle.
The NB921-depressed objects with a $> 2 \sigma$ significance are
labeled from a to g, sorted in order of the $z^{\prime}$ magnitude.
The magenta, green, light blue, and deep blue lines denote
0$\sigma$, 1$\sigma$, 2$\sigma$, and 3$\sigma$ uncertainties
of the color of $z^{\prime} - NB921$, respectively.
The lower direction in this diagram means a depression of
the $NB921$ flux compared to the $z^{\prime}$ flux.
($Right$)
Same as the left panel but shown as a function of
$i^{\prime} - z^{\prime}$ color.
For the data with only upper limited $i^{\prime} - z^{\prime}$ 
colors, an arrow is given.
The NB921-depressed objects with a $> 2 \sigma$ significance are labeled
also in this panel.
The green, blue and magenta lines denote the expected color of model 
galaxies with $EW$(Ly$\alpha$)$_0$ = 65, 130. and 260${\rm \AA}$,
respectively.
\label{fig3}}
\end{figure*}

\end{document}